\documentclass{aa}
\usepackage{graphicx}
\usepackage{booktabs}
\usepackage{txfonts}
\usepackage{mathrsfs}

\begin{document} 

   \title{Constraining differential rotation of Sun-like stars from asteroseismic and starspot rotation periods}

   \author{M.~B.~Nielsen
          \inst{1,2}
          \and
           H.~Schunker
          \inst{2}
          \and
                L.~Gizon
          \inst{2,1}
           \and
                W.~H.~Ball
          \inst{1}
          }

   \institute{Institut f{\"u}r Astrophysik, Georg-August-Universit{\"a}t G{\"o}ttingen, Friedrich-Hund-Platz 1, 37077 G{\"o}ttingen, Germany\\
             \email{nielsenm@mps.mpg.de}
             \and
                                 Max-Planck-Institut f{\"u}r Sonnensystemforschung, Justus-von-Liebig-Weg 3, 37077 G{\"o}ttingen, Germany
             }

   \date{Received XXXXXX XX, 2015; accepted XXXXXX XX, 2015}

  \abstract
        {In previous work, we identified six Sun-like stars observed by \textit{Kepler} with exceptionally clear asteroseismic signatures of rotation. Here, we show that five of these stars exhibit surface variability suitable for measuring rotation. We compare the rotation periods obtained from light-curve variability with those from asteroseismology in order to further constrain differential rotation. The two rotation measurement methods are found to agree within uncertainties, suggesting that radial differential rotation is weak, as is the case for the Sun. Furthermore, we find significant discrepancies between ages from asteroseismology and from three different gyrochronology relations, implying that stellar age estimation is problematic even for Sun-like stars.}

   \keywords{Asteroseismology – Stars: rotation – Stars: Solar-type – Methods: data analysis}
   \titlerunning{Asteroseismic and starspot rotation comparison in Sun-like stars}
   \authorrunning{Nielsen et al.}
   \maketitle

\section{Introduction}
Recently, asteroseismology has become a valuable tool to study the internal rotation of stars. This has been done on a variety of different stars \citep[see, e.g.,][]{Aerts2003,Charpinet2009,Kurtz2014}, including Sun-like stars \citep{Gizon2013,Davies2015}. Asteroseismic measurements of rotation in Sun-like stars have generally been limited to the average internal rotation period. More recently, \citet{Nielsen2014} identified six Sun-like stars where it was possible to measure rotation for independent sets of oscillations modes and found that radial differential rotation is likely to be small. 

From the Sun, we know that different methods for measuring rotation, e.g., spot tracing \citep{DSilva1994} and helioseismology \citep{Schou1998}, show the outer envelope of the Sun rotating differentially. The solar surface rotation period changes by approximately 2.2 days between the equator and $40^{\circ}$ latitude \citep{Snodgrass1990}, while in the radial direction the rotation period changes by ${\sim}1$~day in the outer few percent of the solar radius \citep{Beck2000}. While small on the Sun, differential rotation could be larger on other stars. With data available from the \textit{Kepler} mission \citep{Borucki2010}, it is now possible to compare the measurements of rotation using both asteroseismology and observations of surface features.

Differences between the results of the two methods would imply that differential rotation exists between the regions where the two methods are sensitive; namely, the near-surface layers for asteroseismology, and the currently unknown anchoring depth of active regions in Sun-like stars. However, given the similar scales of the radial and latitudinal differential rotation seen in the Sun, both of these mechanisms may contribute to an observed difference. On the other hand, agreement between the two methods would suggest weak differential rotation. Furthermore, this would also mean that we can measure rotation with asteroseismology in inactive stars that exhibit few or no surface features at all, thus giving us an additional means of calibrating gyrochronology relations \citep[e.g.,][]{Skumanich1972, Barnes2007, Barnes2010}, which so far have primarily relied on measuring the surface variability of active stars.

However, the acoustic modes (or $p$-modes) used to measure rotation in Sun-like stars are damped by surface magnetic activity \citep[][]{Chaplin2011c}. It is therefore difficult to find stars with both a high signal-to-noise oscillation spectrum and coherent signatures of rotation from surface features. \citet{Nielsen2014} identified six Sun-like stars from the \textit{Kepler} catalog with exceptional signal to noise around the p-mode envelope. Five of these stars exhibit rotational variability from surface features. We compare the surface variability periods with those measured from asteroseismology, with the aim of further constraining the near-surface differential rotation of these stars.

\section{Measuring rotation periods}
\begin{figure}
\centering
\includegraphics[width = \columnwidth]{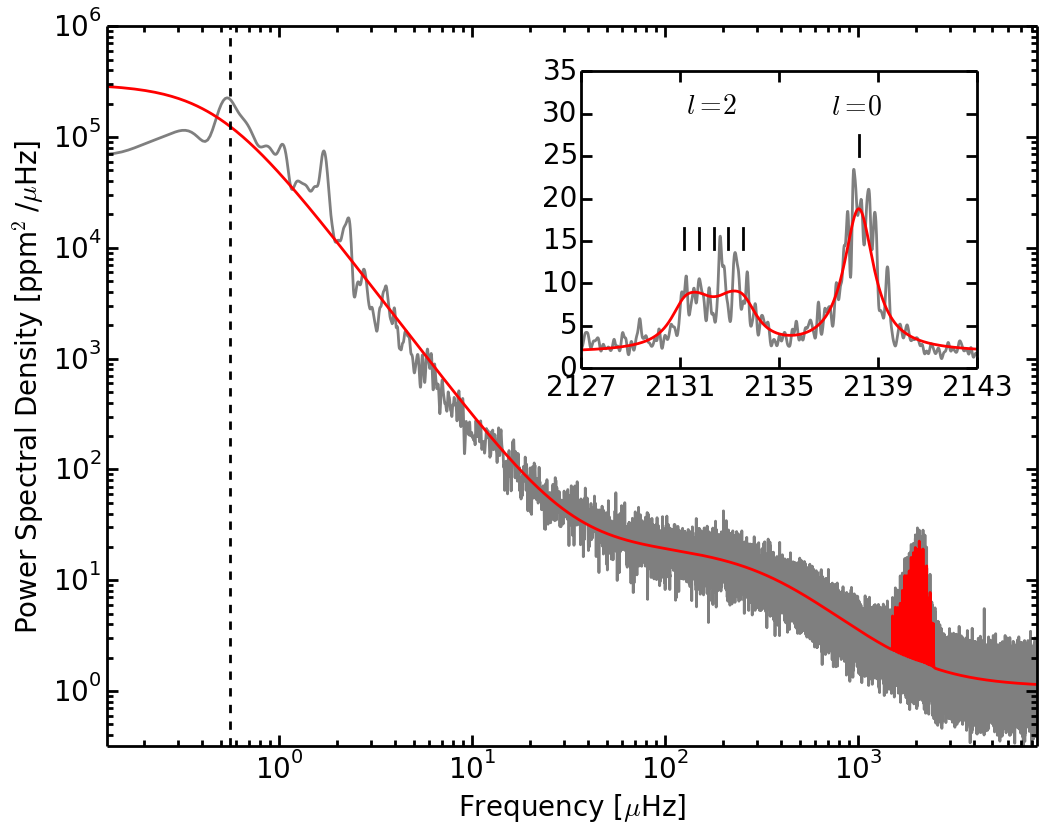}
\caption{Power density spectrum of the short cadence white light photometry time series of \object{KIC005184732}. Gray represents the spectrum smoothed with a $0.1\mu$Hz wide Gaussian kernel, with the best-fit model shown in red. Black dashed indicates the surface variability frequency. The inset shows a zoom of the p-mode envelope at an $l=2,0$ pair, where the black markers indicate the individual $m$-components. The separation between components of the $l=2$ mode is the rotational splitting.}
\label{fig:global_fit}
\end{figure}
We use the ${\sim} 58$~second cadence observations from the Kepler satellite to capture the high frequency acoustic oscillations, and the ${\sim} 29$~minute cadence observations to search for surface variability. The stars we identified were chosen because of their similarity to the Sun with respect to surface gravity and effective temperature, and for having a high signal-to-noise ratio at the p-mode envelope. We use all the available data\footnote{Downloaded from the Mikulski Archive for Space Telescopes, http://archive.stsci.edu/kepler/} for each star, from quarter 1 to 17, spanning approximately four years.

\subsection{Asteroseismic rotation periods}
\label{sec:astero}
The oscillation modes in a star can be described with spherical harmonic functions, characterized by the angular degree $l$ and azimuthal order $m$, and a radial component with order $n$. When the star rotates the modes with azimuthal order $m > 0$ become Doppler shifted, and if this can be measured, the rotation of the star can be inferred. For Sun-like stars, the oscillation modes have a sensitivity weighted toward the surface of the star, and so the asteroseismic measurements predominantly probe the rotation of the star in this region \citep[see, e.g.,][]{Lund2014a}.

To measure this frequency splitting we fit a model to the power spectral density. The modes in a solar-like oscillator are stochastically excited and damped oscillations can be described by a series of Lorentzian profiles, 
\begin{equation}
P(\nu) = \sum\limits_{n} {\sum\limits_{l = 0}^3 {\sum\limits_{m = - l}^l {\frac{{\mathscr{E}_{lm} \left( i \right)S_{nl} }}{{1 + (2/\Gamma_{nlm})^2 \left( {\nu - \nu _{nlm} } \right)^2 }}} } } + B\left( \nu \right) ,
\end{equation}
where $\nu$ is the frequency, and $B(\nu)$ is the background noise.
Here the sums are over the radial orders (typically ${\sim} 8$ values of $n$), the angular degrees $l \leq 3$ and the azimuthal orders $-l \leq m \leq l$. Not all radial orders show clear $l=3$ modes, and so only a few $l=3$ modes are included in the fits, depending on the star. The mode power in a multiplet, $S_{nl}$, is a free parameter. The mode visibility $\mathscr{E}_{lm}(i)$ is a function of the inclination angle $i$ of the rotation axis relative to the line of sight \citep[see][]{Gizon2003}. We assume here that all modes share the same value of $i$. 

The frequencies $\nu_{nlm}$ of the modes reveal the rotation information. For Sun-like stars rotating no more than a few times the solar rotation rate the mode frequencies can be parametrized as $\nu _{nlm} \approx \nu _{nl} + m\delta \nu$, where $\delta \nu$ is the rotational splitting. In our model fit we assume that the stellar interior rotates as a solid body, so that all modes share a common rotational splitting.

The full widths $\Gamma_{nlm}=\Gamma(\nu_{nlm})$ of the Lorentzian profiles depend on the mode lifetimes, which in turn depend on the mode frequencies $\nu_{nlm}$. In the Sun this variation with frequency can be modeled by a low-order polynomial \citep{Stahnthesis}. We therefore opt to parametrize the mode widths as a third-order polynomial in $(\nu-\nu_{\rm max}),$ where $\nu_{\rm max}$ is the frequency at maximum power of the p-mode envelope. Note that only the $l=0$ modes contribute to the fit of this function since they are not broadened or split by rotation. 

The background noise level $B\left( \nu \right)$ is caused by several terms: the very long-term variability (hours to days) from magnetic activity, the variability (tens of minutes) from stellar surface granulation, and the photon noise. For asteroseismology applications, these noise components can be adequately described by two Harvey-like background terms \citep[e.g.,][]{Handberg2011} and a constant term to account for the white photon noise.

We find the best fit with maximum likelihood estimation using a Markov Chain Monte Carlo sampler \citep{Foreman-Mackey2013}. The best-fit values of each parameter are estimated by the median of the corresponding marginalized posterior distribution, and the $16th$ and $84th$ percentile values of the distributions represent the errors. We fit the entire spectrum, including all mode parameters and background terms simultaneously (a so-called global fit), in contrast to \citet{Nielsen2014} who fit the background and separate sets of p-modes individually. A comparison between the global rotational splitting obtained in this work, and the mean of the rotational splittings measured by \citet{Nielsen2014} are shown in Table \ref{tab:results}.

An example spectrum of \object{KIC005184732} is shown in Fig.~\ref{fig:global_fit}. The background terms dominate the low frequency end of the spectrum, while the p-mode envelope appears clearly above the noise level at ${\sim}1800$~$\mu$Hz. The inset shows a part of the p-mode envelope, where the splitting of an $l=2$ mode is clearly visible.

\subsection{Surface variability periods}
The rotation of stars can also be inferred from the periodic variability of their light curves caused by surface features on the stellar disk, such as active regions. For the Sun, active regions are good tracers of the surface rotation of the plasma, to within a few percent \citep{Beck2000}. In other stars this difference could potentially be larger.

\begin{figure}
\centering
\includegraphics[width = \columnwidth]{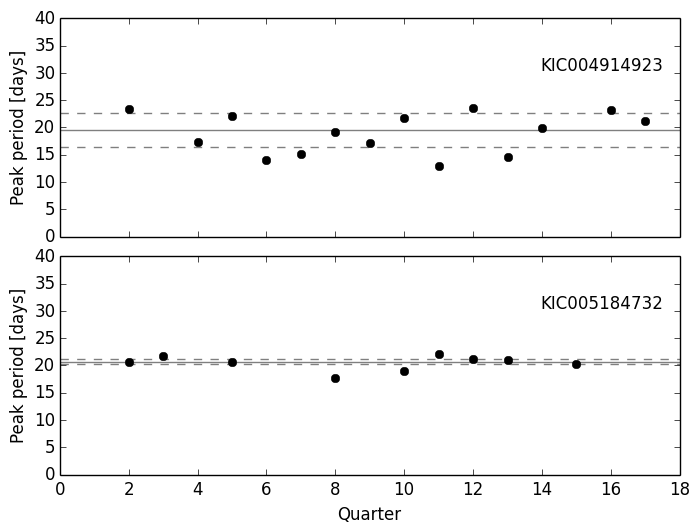}
\caption{Peak periods in the Lomb-Scargle periodogram as a function of observation quarter for \object{KIC004914923} (top) and \object{KIC005184732} (bottom). Solid gray shows the median period, and dashed gray shows the median absolute deviation.}
\label{fig:spot_period}
\end{figure}
From the periodic variation of the stellar light curve it is relatively straightforward to identify the rotation period of the star, and this has been done using automated routines for tens of thousands of stars \citep{Nielsen2013,Reinhold2013,McQuillan2014}. However, these routines typically rely on coherent surface variability over long time periods. The stars studied here do not show simple and regular surface variability and, therefore, they do not appear in these catalogs. Furthermore, the PDC\_MAP pipeline \citep{Smith2012,Stumpe2014} is known to suppress variability on timescales longer than ${\sim}20$ days \citep{Christiansen2013}

We therefore manually reduce the raw pixel data for each star, and search this and the PDC\_MAP reduced data for signs of rotational variability. To reduce the raw data, we use the \textit{kepcotrend} procedure in the PyKe software package\footnote{http://keplerscience.arc.nasa.gov/PyKE.shtml}. The PyKe software uses a series of so-called cotrending basis vectors (CBVs) in an attempt to remove instrumental variability from the raw light curves. The CBVs are computed based on variability common to a large sample of stars on the detector, and so in principle represent the systematic variability. The number of CBVs to use in the reduction is not clear and we therefore compute several sets of reduced light curves using a sequentially increasing number of CBVs (up to six) for each star. We require that the variability appears in light curves reduced with different numbers of CBVs. This minimizes the risk of variability signatures that are caused by the reduction.

To measure the rotation period of each star we used the method of \citet{Nielsen2013}. We compute the Lomb-Scargle periodogram \citep{Lomb1976,Scargle1982} of each quarter and identify the peak of maximum power for periods less than 45 days. The median period of these peaks is used as a first order estimate of the rotation period. We then assume that active regions appear at similar latitudes, and so should have similar periods. This is done by requiring that peaks must lie within 4 median absolute deviations (MAD) from the median. The median and MAD of the remaining peak periods are the measured rotation period and error on the rotation period. Examples are shown in Fig.~\ref{fig:spot_period}. The rotation periods of these stars measured using this method agree with those found by \citet{Garcia2014} \citet{Nascimento2014}.

\begin{table*}
\centering
\caption{List of measured and computed parameters for the five stars.}
\begin{tabular}{lccccc}
\toprule
  & \object{KIC004914923} & \object{KIC005184732} & \object{KIC006116048} & \object{KIC006933899} & \object{KIC010963065} \\
\midrule
  Mass [$\mathrm{M}_{\odot}$] & $1.118\pm0.020$ & $1.205\pm0.025$ & $1.023\pm0.021$  & $1.096\pm0.026$ & $1.062\pm0.021$ \\ [2pt]
  Radius [$\mathrm{R}_{\odot}$] & $1.378\pm0.009$ & $1.342\pm0.010$ & $1.225\pm0.008$ & $1.574\pm0.025$ & $1.220\pm0.009$ \\[2pt]
  $\mathrm{X_{c}}$ & 0.000 &  0.216 & 0.036 &  0.000 & 0.108 \\[2pt]
  $\tau$ [days]        & $16.94$ & $23.30$ & $34.34$ & $28.25$ & $33.03$ \\[2pt]
  
  $\mathrm{T_{eff}}$[K] & $5880\pm70$ & $5865\pm70$ &  $5990\pm70$ & $5870\pm70$ &  $6090\pm70$ \\[2pt]

  $P_\mathrm{S}$ [days]& $19.49\pm{3.12}$ & $20.69\pm{0.50}$ & $17.96\pm{2.11}$ & $31.63\pm{1.43}$ & $12.27\pm{0.32}$ \\[2pt]
  $P_\mathrm{A}$ [days]& $17.98_{-2.27}^{+3.17}$ & $19.44_{-2.13}^{+1.63}$ & $17.61_{-1.31}^{+0.95}$ & $29.92_{-6.76}^{+4.90}$ & $12.01_{-1.09}^{+1.42}$ \\[2pt]
  $P_\mathrm{N}$ [days]& $22.17\pm3.14$ & $18.00\pm1.76$ & $16.34\pm0.89$ & $28.65\pm5.53$ & $14.45\pm1.43$ \\[2pt]
  
  Seismic age [Gyr]    & $6.23\pm0.36$ & $4.39\pm0.13$ & $5.70\pm0.21$ & $6.57\pm0.30$ & $4.18\pm0.19$ \\[2pt]
  \midrule
    Gyrochronology ages [Gyr]     & & & & & \\
  \citet{Barnes2010}       & $2.68_{-0.62}^{+0.95}$ & $2.72_{-0.54}^{+0.47}$ & $3.09_{-0.44}^{+0.36}$ & $5.58_{-1.95}^{+2.42}$ & $1.97_{-0.34}^{+0.49}$ \\ [3pt]
  \citet{Mamajek2008}      & $3.56_{-1.14}^{+1.93}$ & $2.36_{-0.53}^{+0.63}$ & $3.36_{-0.82}^{+1.14}$ & $8.57_{-3.08}^{+3.73}$ & $4.38_{-1.74}^{+3.77}$ \\[3pt]
  \citet{Garcia2014}       & $3.25_{-0.95}^{+1.52}$ & $3.65_{-0.95}^{+1.37}$ & $3.05_{-0.67}^{+0.92}$ & $7.83_{-2.84}^{+4.78}$ & $1.49_{-0.34}^{+0.51}$ \\[3pt]
\midrule 

\end{tabular}
\tablefoot{The stellar masses and radii, core hydrogen fraction $\mathrm{X_{c}}$, and convective turnover time $\tau$, as well as the seismic ages are computed from the best-fit models as described in Section \ref{sec:gyro}. The effective temperatures are adapted from \citet{Bruntt2012}. The measured surface variability rotation periods $P_\mathrm{S}$ and asteroseismic rotation periods $P_\mathrm{A}$ are also listed. These periods are compared to the averaged asteroseismic rotation periods $P_\mathrm{N}$ from \citet{Nielsen2014}. The $P_\mathrm{N}$ values are averages over several sets of oscillation modes. The gyrochronology ages are calculated from \citet{Barnes2010}, \citet{Mamajek2008}, and \citet{Garcia2014}.}
\label{tab:results}
\end{table*}

\section{Comparing asteroseismic and surface variability periods}
\begin{figure}
\centering
\includegraphics[width = \columnwidth]{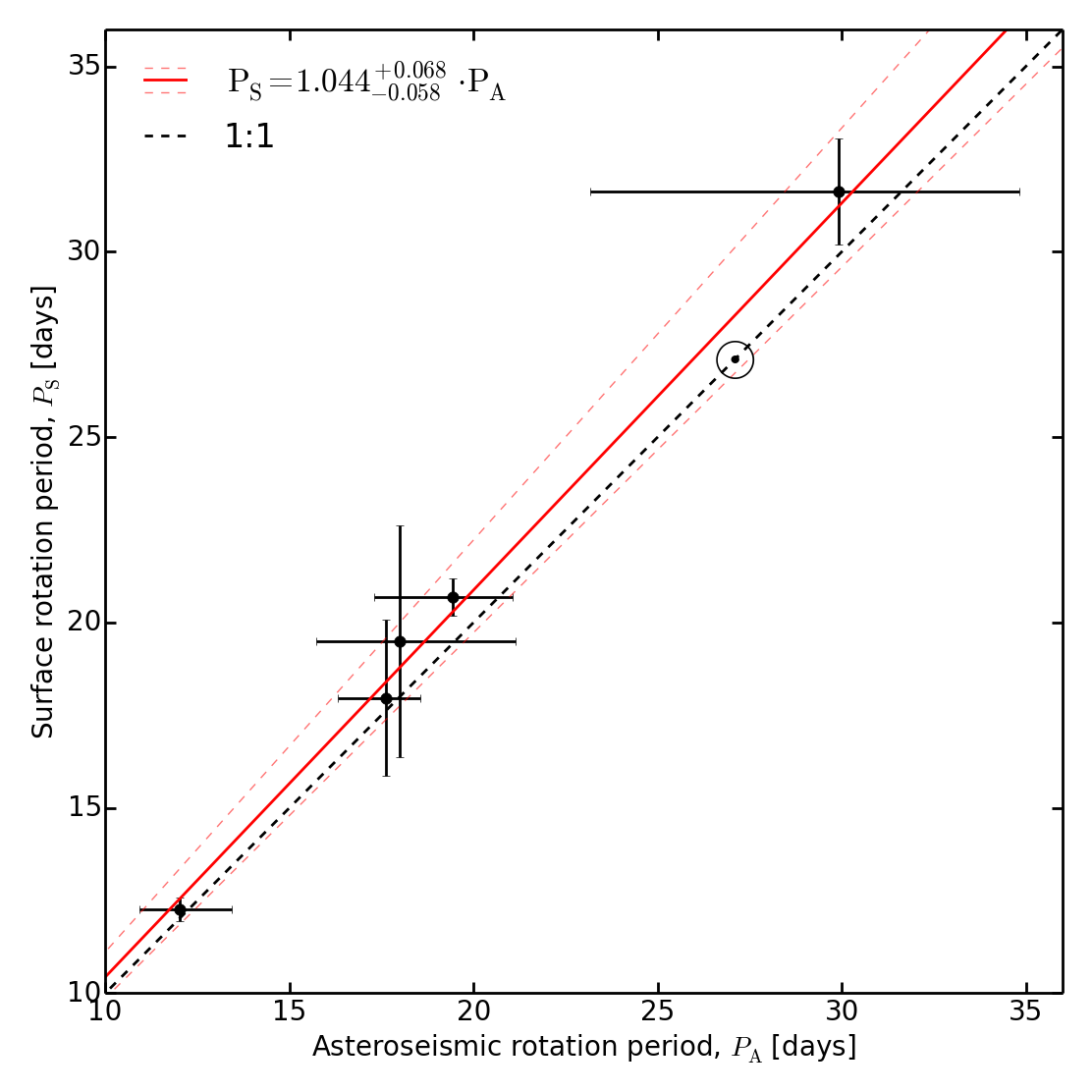}
\caption{Surface variability period $P_\mathrm{S}$ as a function of asteroseismic period $P_\mathrm{A}$. The 1:1 line is shown with a black dashed line. The best-fit value and errors to the slope are shown with red solid and dashed lines, respectively. For comparison the Carrington period of 27.3~days is also shown (this is not included in the fit).}
\label{fig:results}
\end{figure}

In Table~\ref{tab:results} we list the periods for each analysis. These values are also shown in Fig. \ref{fig:results}. The asteroseismic and surface variability periods agree very well for all the stars analyzed here. However, the asteroseismic rotation period measurements are systematically lower than those from surface variability. As a first order significance estimate of this difference we fit a linear function $P_\mathrm{S} = a P_\mathrm{A}$ to the measured values. We find the best fit by maximum likelihood estimation, using the marginalized posterior distributions for the asteroseismic measurements, and assume Gaussian posterior distributions for the surface variability periods. The best-fit solution returns $a = 1.044_{-0.058}^{+0.068}$, showing that the difference between the two measurement methods is insignificant (see Fig. \ref{fig:results}). As a consistency check we perform an identical fit to the average rotation periods $P_\mathrm{N}$ from \citet{Nielsen2014}, assuming Gaussian errors (see Table \ref{tab:results}). We find a slope of $1.006^{+0.066}_{-0.060}$, in good agreement with the slope of $P_\mathrm{A}$. However, this does not account for the non-Gaussian form of the marginalized posteriors of the fits performed in that work.

\section{Gyrochronology}
\label{sec:gyro}
\begin{figure}
\centering
\includegraphics[width = \columnwidth]{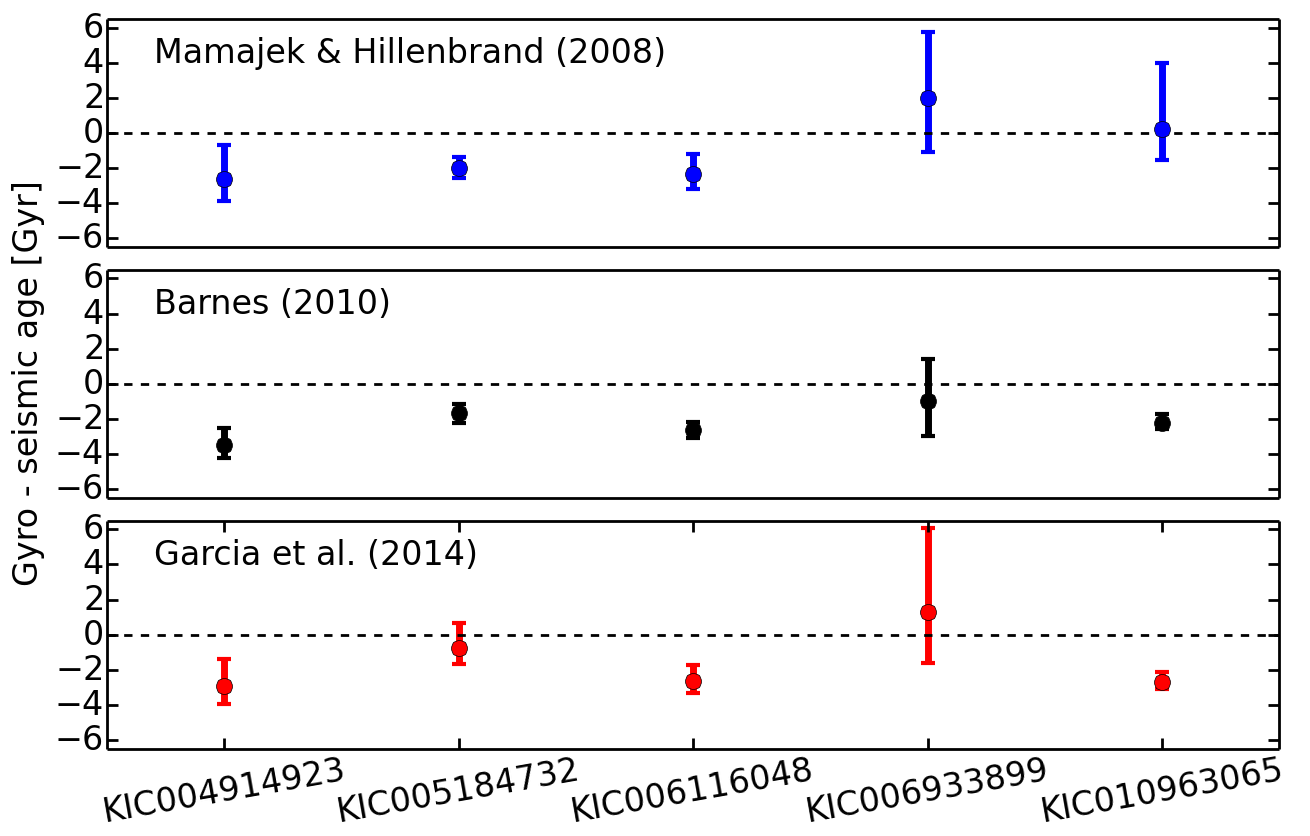}
\caption{Differences between gyrochronology ages and asteroseismic ages for three different gyrochronology relations. The relations are \citet{Mamajek2008}, \citet{Barnes2010}, and \citet{Garcia2014}.}
\label{fig:avp}
\end{figure}

Here we estimate the stellar ages from three different gyrochronology relations provided by \citet{Mamajek2008}, \citet{Barnes2010}, and \citet{Garcia2014}. These relations represent distinct methods of estimating gyrochronology ages. In all these relations, we use the asteroseismic rotation periods from Sect.~\ref{sec:astero}.

The type of gyro-relation used by \citet{Mamajek2008}, first proposed by \citet{Barnes2007}, assumes that the rotation period of a star varies as powers of age and mass, where the $B-V$ color is used as a proxy for the latter. This relation is calibrated to young cluster stars with ages $\lesssim 625$~Gyr, with only the Sun representing the older stellar populations. For the stars studied here we use $B-V$ values from \citet{Hoeg2000}, using extinction values from the KIC catalog \citep{Brown2011}. 

The relation by \citet{Garcia2014} is calibrated to stars with asteroseismically determined ages from \citet{Mathur2012} and \citet{Metcalfe2014}. However, this relation does not include a mass dependence since the calibrators are all similar in mass (${\sim}1\mathrm{M}_\odot$), and so this relation only depends on the rotation period of the star.

Lastly, the relation by \citet{Barnes2010} incorporates the effects of a rapid spin-down phase, experienced by very young stars (${\sim}100$~Myr), into the prediction of the stellar age. This relation uses convective turnover time $\tau$ as a proxy for mass, and is also calibrated using young cluster stars, with the Sun as the only anchoring point for older stars.

We determine the convective turnover times from stellar models fit to the stellar oscillation frequencies (see online material in \citet{Nielsen2014}). These stellar models are produced using the Modules for Experiments in Stellar Astrophysics \citep[MESA,][]{Paxton2011, Paxton2013}, with the same input physics as in \citet{Ball2014}. Initial fits were determined using the SEEK method \citep{Quirion2010} to compare a grid of models to observed spectroscopic and global asteroseismic parameters. The models cover the mass range $0.60$--$1.60\mathrm{M_\odot}$ and initial metal abundances $Z$ in the range $0.001$--$0.040$. The helium abundance was presumed to follow the enrichment law $Y=0.245+1.45Z$ and the mixing length parameter was fixed at $\alpha_\odot=1.908$.
%\LEt{Please introduce this acronym and ADIPLS below. Please check acronyms below as well (K2 \& PLATO)}

As initial guesses, we used the median values of the grid-based fit and also generated ten random, uniformly-distributed realizations of the parameters. The 11 initial guesses were optimized in five parameters (age, mass, metallicity, helium abundance, and mixing length) using a downhill simplex \citep{Nelder1965} to match spectroscopic data \citep{Bruntt2012} and individual oscillation frequencies, computed with ADIPLS \citep{Adipls} and corrected according to the cubic correction by \citet{Ball2014}. As in \citet{Ball2014}, the seismic and nonseismic observations are weighted only by their respective measurement uncertainties. Best-fit parameters and uncertainties are estimated from ellipses bounding surfaces of constant $\chi^2$ for all the models determined during the optimizations, corresponding to a total of about 3000 models for each star.

The best-fit models yield local convective velocities and we use these to compute $\tau$ as in \citet{Kim1996}, these are listed in Table \ref{tab:results}. From these values of $\tau$ and the asteroseismic rotation periods from section \ref{sec:astero}, we compute the gyrochronology ages using Eq.~32 from \citet{Barnes2010} and the suggested calibration values therein.

In addition to providing values of $\tau,$ the stellar models also produce estimates of the stellar ages. In Fig. \ref{fig:avp} these ages are compared with those derived from the three gyrochronology relations. It is immediately obvious that the gyro-relations are in general not consistent with the seismic ages. For some stars a particular gyro-relation may agree well with the seismic age, but for other stars that same relation deviates significantly, as seen, e.g., for the ages from \citet{Garcia2014} and \citet{Mamajek2008} relations for \object{KIC005184732} and \object{KIC010963065}. However, it appears that the gyro-relations tend to underestimate the stellar ages relative to the seismic values.

For the star \object{KIC006933963,} the gyro-relations predict an age older than the stellar models, which is likely because this star is a subgiant, with a slightly contracted core and expanded envelope. Thus, by angular momentum conservation one expects the surface and near-surface regions to have slowed down faster than the gyro-relations predict, leading to an overestimated age. \object{KIC004914923} also appears to be a subgiant based on its core hydrogen content (see Table \ref{tab:results}), but close inspection of the best-fit stellar model shows that this star has only just moved off the main sequence, and so may still adhere to these gyro-relations. 

It should be noted that the errors on the asteroseismic ages are internal errors from the fit. However, even conservatively assuming a precision of ${\sim}20\%$ on the asteroseismic ages \citep{Aerts2015}, the two methods of age estimation are still in poor agreement. 

\section{Conclusions}
We measured the rotation periods of five Sun-like stars, using asteroseismology and surface variability. The measurements show that the asteroseismic rotation periods are ${\sim}4\%$ lower than those from surface variability, but this was not found to be a statistically significant difference given the measurement uncertainties. Thus there is no evidence for any radial (or latitudinal) differential rotation between the regions that the two methods are sensitive to.

Previous efforts to calibrate gyrochronology relations have focused on measuring rotation from surface variability of cluster stars. Given that rotation periods can be reliably determined by asteroseismology compared to the surface variability, gyrochronology may also be applied to very inactive stars for which surface variability rotation periods are not available.

Our comparison of three different gyrochronology relations from the literature found that these ages differed by up to several Gyr from the seismic ages. In general, the gyro-relations seemed to underestimate the stellar ages relative to the asteroseismic ages. It also demonstrated that the gyro-relations are not internally consistent.

When considering these discrepancies it is important to remember the calibrations used for each method. The \citet{Barnes2010} and \citet{Mamajek2008} relations are both calibrated using ages and rotation periods of the Sun and cluster stars, and so may not accurately predict ages of stars similar to or older than the Sun \citep[e.g.,][]{Meibom2015}. On the other hand, asteroseismic ages from stellar models are only calibrated to the Sun, and in addition are dependent on the input physics and assumptions of the stellar models \citep{Lebreton2014}. The relation by \citet{Garcia2014} lacks a mass dependent term, in contrast to the two other tested relations. However, this was not expected to have a remarkable effect since the stars in our sample were approximately $1\mathrm{M}_{\odot}$ stars, similar to the calibrators used by \citet{Garcia2014}. Despite this, there is still a discrepancy of several Gyr between the ages derived from this relation and our seismic ages for some of the stars.

Tighter asteroseismic constraints for stellar modeling will become available when calibrations are obtained from cluster stars to be observed by space missions like K2 \citep{Chaplin2015} and the Planetary Transits and Oscillations of stars (PLATO) mission \citep{Rauer2014}. This in turn will allow us to reliably use asteroseismic rotation periods to further refine gyrochronology relations for magnetically inactive field stars.

\begin{acknowledgements}
The authors acknowledge research funding by Deutsche Forschungsgemeinschaft (DFG) under grant SFB 963/1 ``Astrophysical flow instabilities and turbulence'' (Project A18). L.G. acknowledges support from the Center for Space Science at NYU Abu Dhabi Institute under grant 73-71210-ADHPG-G1502. This paper includes data collected by the \textit{Kepler} mission. Funding for the \textit{Kepler} mission is provided by the NASA Science Mission directorate. Data presented in this paper were obtained from the Mikulski Archive for Space Telescopes (MAST). STScI is operated by the Association of Universities for Research in Astronomy, Inc., under NASA contract NAS5-26555. Support for MAST for non-HST data is provided by the NASA Office of Space Science via grant NNX09AF08G and by other grants and contracts. 
\end{acknowledgements}

\bibliographystyle{aa}
\bibliography{main}

\end{document}